\documentclass[12pt,a4paper]{amsproc}

\usepackage[margin=20mm]{geometry}
\usepackage{amsmath,amssymb,amsthm}
\usepackage{mathrsfs}
\usepackage{diagbox}
\usepackage{graphicx}
\usepackage{color}

\newtheorem{theorem}{Theorem}
\newtheorem{corollary}{Corollary}
\newtheorem{lemma}{Lemma}

\theoremstyle{definition}
\newtheorem{assumption}{Assumption}
\newtheorem{remark}{Remark}

\numberwithin{equation}{section} 

\DeclareMathOperator*{\argmin}{arg\,min}
\DeclareMathOperator*{\Var}{Var}

\def\R{\mathbb{R}}
\def\1{\mathbf{1}}
\def\ve{\varepsilon}
\def\hve{\hat\ve}
\def\sj{\sum_{j=1}^n}

\def\avj{\frac1n \sj}

\def\opn{o_p(n^{-1/2})}
\def\op{o_p(1)}
\def\GG{\mathscr{G}}
\def\FF{\mathscr{F}}
\def\RR{\mathscr{R}}
\def\BB{\mathscr{B}}

\def\Tspace{\mathscr{T}}
\def\Lspace{\mathcal{L}}
\def\Vspace{\mathcal{V}}
\def\Sspace{\mathcal{S}}
\def\rhat{\hat{r}}
\def\ahat{\hat{a}}
\def\Femp{\mathbb{F}}
\def\Fhat{\hat{\Femp}}
\def\Ftilde{\tilde{\Femp}}

\allowdisplaybreaks

\begin{document}

\title[Efficiently estimating the error distribution]{
  Efficiently estimating the error distribution in nonparametric
  regression with responses missing at random}

\author[J.\ Chown and U.U.\ M\"uller]{Justin
  Chown$^{\text{a}\ast}$  and Ursula U. M\"uller$^{\text{b}\ast}$
}
\thanks{{\em $^{\ast}$ Correspondence may be addressed to either author: \\
  \indent $^{\text{a}}$ Fakult\"at f\"ur Mathematik, Lehrstuhl f\"ur
    Stochastik, 44780 Bochum, DE \\
  \indent Email: justin.chown@ruhr-uni-bochum.de \\
  \indent $^{\text{b}}$ Department of Statistics, Texas A\&M University,
    College Station, TX 77843-3143, USA \\
  \indent Email: uschi@stat.tamu.edu
}}

\begin{abstract}
This article considers nonparametric regression models with
multivariate covariates and with responses missing at random. We
estimate the regression function with a local polynomial smoother. The
residual-based empirical distribution function that only uses complete
cases, i.e.\ residuals that can actually be constructed from the data,
is shown to be efficient in the sense of H\'ajek and Le Cam. In the
proofs we derive, more generally, the efficient influence function for
estimating an arbitrary linear functional of the error distribution;
this covers the distribution function as a special case. We also show
that the complete case residual-based empirical distribution function
admits a functional central limit theorem. The article concludes with
a small simulation study investigating the performance of the complete
case residual-based empirical distribution function.
\end{abstract}
\bigskip

\maketitle

\noindent {\em keywords:}
efficient estimator, empirical distribution function, local polynomial
smoother, martingale transform, missing at random, nonparametric
regression, test for normal errors, transfer principle
\bigskip



\section{Introduction and main result}
\label{intro}
An important tool for making decisions about goodness-of-fit and
lack-of-fit is the residual-based empirical distribution
function. This has been studied in many articles. Stute (1997) and
Khmaladze and Koul (2004, 2009), for example, test parametric
hypo\-theses about the regression function in nonparametric
models. Neumeyer and Van Keilegom (2010) study additivity tests in
heteroskedastic nonparametric regression. M\"uller, Schick and
Wefelmeyer (2012) test for normal errors.

In this article we study the nonparametric regression model
\begin{equation} \label{modeleq}
Y = r(X) + \ve,
\end{equation}
with the error $\ve$ independent of the covariate vector
$X$. Nonparametric models are particularly useful for residual-based
inference because residuals constructed from them are usually
consistent. We are interested in the case where responses $Y$ are
missing, i.e.\ we observe the sample $(X_1, \delta _1 Y_1, \delta_1),
\dots, (X_n, \delta _n Y_n, \delta _n)$, where $\delta$ is an
indicator variable which equals one, if $Y$ is observed, and zero,
otherwise. In practical applications, most datasets contain missing
responses. It is important to choose appropriate statistical methods
that ensure conclusions are not biased.

We make the assumption that responses are {\em missing at random}
(MAR). This means that the probability that $Y$ is observed depends
only on the covariates,
\begin{equation*}
P(\delta =1|X,Y) = P(\delta =1|X) = \pi(X).
\end{equation*}
We will refer to the model with responses missing at random as the
{\em MAR model}. MAR is a common assumption and is reasonable in many
situations (see Little \& Rubin, 2002, Chapter 1). As an example,
consider missing responses to a survey question about income. If
additional data ($X$) about medical conditions were available, we
might see that the response probabilities ($\pi$) are smaller for
subjects diagnosed with depression. In this case the missing mechanism
is ignorable since $\pi$ depends only on fully observed data $X$,
i.e.\ it can be estimated from the data. More examples of missing data
can be found in Tsiatis (2006), in Liang, Wang and Carroll (2007), in
Molenberghs and Kenward (2007), and in Efromovich (2011a, 2011b).

We show in this article that the residual-based empirical distribution
function $\Fhat_c$ given in equation \eqref{FhatCC} below is an
efficient estimator of the unknown error distribution function
$F$. This estimator uses only the complete data pairs $(X,Y)$, i.e.\
the available residuals $\hve_{j,c} = Y_j - \rhat_c(X_j)$, where
$\rhat_c$ is a suitable complete case estimator of the regression
function.

Demonstrating this requires two steps. First we show that $\Fhat_c$
satisfies the uniform stochastic expansion
\begin{equation} \label{expN}
\sup_{t \in \R} \bigg|
 \Fhat_c(t) - \frac{1}{N} \sj \delta_j \1\big[\ve_j \leq t\big]
 - f(t) \frac{1}{N} \sj \delta_j \ve_j \bigg| = \opn.
\end{equation}
Here $f$ is the error density and $N = \sj \delta _j$ is the number of
complete cases. Then we show that an estimator of $F$ that admits this
expansion is asymptotically efficient in the sense of H\'ajek and Le
Cam. This follows from the arguments in Section \ref{eff}, where we
derive, more generally, the efficient influence function for
estimating an arbitrary linear functional $E[h(\ve)]$, which covers
$F(t) = E[1(\ve \le t)]$ as a special case. We conclude that an
estimator $\Fhat_c$ with expansion \eqref{expN} is indeed efficient
for $F$.

We handle part of the proof that \eqref{expN} holds by using the {\em
  transfer principle} for complete case statistics in Koul, Schick and
M\"uller (2012). This principle makes it possible to adapt results for
the model where all data are fully observed, the {\em full model}, to
missing data models. In particular, we can use the complete case
version $\rhat_c$ of an estimator $\rhat$ in the fully observed data
model (i.e.\ all indicators $\delta_j$ are equal to one). M\"uller,
Schick and Wefelmeyer (2009) obtain expansion \eqref{expN} for the full
model using a local polynomial smoother to estimate the regression
function $r$, and these authors derive useful results the local
polynomial estimator of the regression function that are suitable to
the missing data model considered here. See also Neumeyer and Van
Keilegom (2010), who consider heteroskedastic nonparametric
regression.

In order to summarize the main result by M\"uller, Schick and
Wefelmeyer (2009) (Theorem \ref{MSWthm1} below), we introduce some
notation. Let $i = (i_1,\dots,i_m)$ be a multi-index and write $I(k)$
for the set of multi-indices that satisfy $i_1 + \dots + i_m \le
k$. M\"uller, Schick and Wefelmeyer (2009) estimate $r$ by a local
polynomial smoother $\rhat$ of degree $d$. It is defined as the
component $\hat \beta_0$ corresponding to the multi-index
$0=(0,\dots,0)$ of a minimizer
\begin{equation*}
\hat \beta = \argmin_{\beta=(\beta_i)_{i\in I(d)}} \sj
 \bigg\{ Y_j - \sum_{i\in I(d)} \beta_i
 \psi_i\bigg(\frac{X_j - x}{c_n}\bigg) \bigg\}^2
 w\bigg(\frac{X_j - x}{c_n}\bigg),
\end{equation*}
where 
\begin{equation*}
\psi_i(x) = \frac{x_1^{i_1}}{i_1!}
 \cdots\frac{x_m^{i_m}}{i_m!},
 \qquad x=(x_1,\dots,x_m)\in \R^m,
\end{equation*}
$w(x)= w_1(x_1)\cdots w_m(x_m)$ is a product of densities, and
$\{c_n\}_{n \geq 1}$ is a bandwidth sequence.

The estimator $\rhat$ permits the desired expansion, if the
assumptions of Theorem \ref{MSWthm1} (below) are satisfied. This
requires, in particular, the regression function $r$ belongs to the
H\"older space $H(d, \gamma)$, i.e.\ it has continuous partial
derivatives of order $d$ (or higher), and that the partial derivatives
of order $d$ are H\"older with exponent $\gamma$. The choice of the
degree $d$ of the local polynomial smoother will also depend on
smoothness and moment conditions on the error density, and on the
dimension of the covariate vector. In our simulation study in Section
\ref{sims}, we consider an infinitely differentiable regression
function $r$ and a one-dimensional covariate $X$, which allows us to
use a locally linear smoother. Theorem 1 from M\"uller, Schick and
Wefelmeyer (2009) is proved under the following assumption on the
covariate distribution:
\begin{assumption} \label{assumpG}
The covariate vector $X$ is quasi-uniform on the cube $[0,\,1]^m$,
i.e.\ $X$ has a density which is bounded and bounded away from zero on
$[0,\,1]^m$.
\end{assumption}

\begin{theorem}[Theorem 1 of M\"uller, Schick and Wefelmeyer, 2009]
\label{MSWthm1}
Let Assumption \ref{assumpG} be satisfied. Suppose that the regression
function $r$ belongs to $H(d,\gamma)$ with $s = d + \gamma >
3m/2$. Let that the error variable have mean zero, a finite moment of
order $\zeta > 4s/(2s - m)$ and a density $f$ that is H\"older with
exponent $\xi > m/(2s - m)$. Consider the estimator $\rhat$ from above
with densities $w_1,\dots,w_m$ that are $(m+2)$-times continuously
differentiable with compact support $[-1,\,1]$. Finally, let the
bandwidth sequence satisfy $c_n \sim (n \log n)^{-1/(2s)}$. Then,
writing $\hve_j = Y_j - \hat r(X_j)$,
\begin{equation*}
\sup_{t\in \R} \bigg| \avj \Big\{
 \1\big[\hve_j \leq t\big] - \1\big[\ve_j \leq t\big]
 - \ve_j f(t) \Big\} \bigg|
 = \opn.
\end{equation*}
\end{theorem} 

We can apply the transfer principle for asymptotically linear
statistics given by Koul, M\"uller and Schick (2012) to adapt the
results from Theorem \ref{MSWthm1} for the MAR model as follows. The
complete case estimator for $F(t)$ is given by
\begin{equation} \label{FhatCC}
\Fhat_c(t) = \frac{1}{N}\sj \delta_j\1\big[\hve_{j,c} \leq t\big]
 = \frac{1}{N}\sj \delta_j\1\big[Y_j - \hat r_c(X_j) \leq t\big],
\end{equation}
where $\rhat_c$ is the complete case version of $\rhat$, i.e.\
$\rhat_c$ is given by the component $\hat \beta_{c0}$ of a minimizer 
\begin{equation} \label{locpolyCC}
\hat \beta_c = \argmin_{\beta=(\beta_i)_{i\in I(d)}} \sj
 \delta_j \bigg\{ Y_j - \sum_{i \in I(d)}
 \beta_i \psi_i\bigg(\frac{X_j - x}{c_n}\bigg) \Big\}^2
 w\bigg(\frac{X_j - x}{c_n}\bigg).
\end{equation}
Using the transfer principle requires the conditional distribution of
$(X,Y)$ given $\delta =1$ to meet the assumptions on the
(unconditional) joint distribution of $(X,Y)$ from Theorem
\ref{MSWthm1}. In our case, it is easy to see how this requirement
affects only the covariate distribution $G$: the MAR assumption
combined with the independence of $X$ and $\ve$ yield that $\ve$ and
$(X,\delta)$ are independent. This implies the parameters $f$ and $r$
stay the same when switching from the unconditional to the conditional
distribution. In particular, the complete case statistic $\Fhat_c(t)$
is a consistent estimator for $F(t)$ in the MAR model (since $F$
remains unchanged). Hence, we can keep all but one of our assumptions:
only Assumption \ref{assumpG} must be restated.
\begin{assumption} \label{assumpG1}
The conditional distribution of the covariate vector $X$ given
$\delta=1$ is quasi-uniform on the cube $[0,1]^m$, i.e.\ it has a
density which is bounded and bounded away from zero on $[0,1]^m$.
\end{assumption}

The transfer principle implies the complete case version of the
estimator from Theorem \ref{MSWthm1} has the corresponding expansion
\eqref{expN}. This expansion is equivalent to 
\begin{equation*}
\sup_{t \in \R} \bigg| \avj \frac{\delta _j}{E\delta} \bigg\{
 \1\big[\hve_{j,c} \leq t\big] - \1\big[\ve_j \leq t\big]
 - \ve_j f(t) \bigg\} \bigg|
 = \opn.
\end{equation*}
Hence, we have, uniformly in $t \in \R$, 
\begin{equation*}
\Fhat_c(t) = \avj \frac{\delta_j}{E\delta}
 \1\big[\hve_{j,c} \leq t\big]  + \opn
 = F(t) + \avj  b(\delta_j, \ve_j, t) + \opn,
\end{equation*}
where $b(\delta,\ve,t) = (\delta/E\delta)\{\1[\ve \leq t] - F(t) + \ve
f(t)\}$ is the influence function. This is indeed the {\em efficient
  influence function} for estimating $F(t)$: see Corollary
\ref{corFhatEffExpan} in Section \ref{eff}. This brings us to the main
result of this paper.
\begin{theorem} \label{thmFhatCCExpan}
Consider the nonparametric regression model with responses missing at
random. Suppose the assumptions of Theorem \ref{MSWthm1} are
satisfied, with Assumption \ref{assumpG1} in place of Assumption
\ref{assumpG}. Then the {\em complete case} estimator $\Fhat_c$ of the
error distribution function $F$ satisfies the stochastic expansion
\eqref{expN},
\begin{equation*}
\sup_{t \in \R} \bigg| \frac1N \sj \delta _j \Big\{
 \1\big[\hve_{j,c} \leq t\big] - \1\big[\ve_j \leq t\big]
 - \ve_j f(t) \Big\} \bigg|
 = \opn.
\end{equation*} 
If the error density $f$ furthermore fulfills Assumption
\ref{assumpF}, stated in Section \ref{eff}, then $\Fhat_c$ is
asymptotically efficient in the sense of H\'ajek and Le Cam for
estimating $F$, with influence function
\begin{equation*}
b(\delta,\ve,t) = \frac{\delta}{E\delta} \Big\{
 \1\big[\ve \leq t\big] - F(t) + \ve f(t) \Big\}.
\end{equation*}
\end{theorem}

\begin{remark} \label{remNoTTPrin}
If the transfer principle were not available, the expansion in Theorem
\ref{thmFhatCCExpan} could be derived by mimicking the (rather
elaborate) proofs of Lemma 1 in M\"uller, Schick and Wefelmeyer (2009)
and of Theorem 2.2 in M\"uller, Schick and Wefelmeyer (2007), who
estimate the error distribution in a general semiparametric regression
model. The arguments are essentially the same -- what is new now is
the presence of indicators.

Analogously to M\"uller, Schick and Wefelmeyer (2009), derive an
approximation $\ahat_c(x)$ of the difference $\hat r_c(x) - r(x)$,
\begin{equation} \label{approx}
\sup_{x \in [0,\,1]^m} \big|
 \rhat_c(x) - r(x) - \ahat_c(x) \big|
 = \opn;
\end{equation}
see equation (1.4) in that paper. Note, the events $\{\hve_{j,c} \leq
t\}$ and $\{\ve_j \leq t + \rhat_c(x) - r(x)\}$ are
equivalent. Combining this fact and \eqref{approx} with replacing the
two empirical distribution functions $\Fhat_c$ and $N^{-1} \sj
\delta_j \1[\ve_j \leq t]$ in the proof of Theorem 2.2 in M\"uller,
Schick and Wefelmeyer (2007) yields
\begin{equation*}
\sup_{t \in \R} \bigg| \avj \frac{\delta _j}{E\delta} \Big\{
 \1\big[\hve_{j,c} \leq t\big] - \1\big[\ve_j \leq t\big]
 - F_{\ahat_c}(t) - F(t) \Big\} \bigg|
 = \opn,
\end{equation*}
writing
\begin{equation*}
F_{\ahat_c}(t) = E\bigg[\frac{\delta _j}{E\delta}
 \1\big[\ve \leq t + \ahat_c(X)\big] \bigg]
 = E\Big[\1\big[\ve \leq t + \ahat_c(X)\big]\,\Big|\,\delta=1\Big]
 = \int_{[0,\,1]^m} F\big(t + \ahat(x)\big) \, G_1(dx).
\end{equation*}
Here $G_1$ denotes the conditional distribution of $X$ given $\delta =
1$. A Taylor expansion applied to the difference $F_{\hat a_c}(t) -
F(t)$ in the above expansion gives
\begin{equation*}
\sup_{t \in \R} \bigg| \avj \frac{\delta _j}{E\delta} \bigg\{
 \1\big[\hve_{j,c} \leq t\big] - \1\big[\ve_j \le t\big]
 - f(t) \int_{[0,\,1]^m} \ahat_c(x)\,G_1(dx) \bigg\} \bigg|
 = \opn.
\end{equation*}
The desired expansion now follows from this combined with
\begin{equation*}
\int_{[0,\,1]^m} \ahat_c(x)\,G_1(dx)
 = \avj \frac{\delta_j}{E\delta} \ve_j + \opn.
\end{equation*}
The last approximation is the complete case version of equation (1.3)
in M\"uller, Schick and Wefelmeyer (2009). It can be verified by
inspecting the proof of Lemma 1 in that paper, where properties of
locally polynomial smoothers are derived. Keep in mind that our
estimators are constructed from the complete cases (equation
\eqref{locpolyCC} above), which explains the indicators in the above
formula.
\end{remark}

Note, the uniform expansion \eqref{expN} implies $\Fhat_c$ satisfies a
functional central limit theorem, and the efficiency property of the
estimator $\Fhat_c$ guarantees that competing estimators will not be
able to outperform it in large samples. This includes estimators based
on imputations that attempt to replace the missing responses. The
article is organized as follows. We provide the efficient influence
function for estimating linear functionals of the error distribution
function $F$ in Section \ref{eff}, and we specialize these results to
estimators of $F$. In Section \ref{sims}, we illustrate this result
with simulations for two examples. The first example demonstrates the
efficiency property of the complete case estimator $\Fhat_c$ by
comparing it with a `tuned' estimator using an imputation technique
that is in the spirit of Gonz\'alez-Manteiga and P\'erez-Gonz\'alez
(2006). For our second example, we perform simulations similar to
those in M\"uller, Schick and Wefelmeyer (2012), who use a martingale
transform approach to test for normal errors in the full model. The
test statistics involve the estimators from the first example.


\section{Efficiency}
\label{eff}
In this section we provide the {\em efficient influence function} for
estimating the linear functional $E[h(\ve)]$ using observations
$(X_i,\delta_i  Y_i,\delta_i )$, $i = 1, \ldots, n$. We first follow
the arguments of M\"uller, Schick and Wefelmeyer (2006), who study
efficient estimation of general differentiable functionals with data
of the above form. We summarize their main arguments and refer to that
paper for more details. We then focus on the functional $E[h(\ve)]$,
which M\"uller, Schick and Wefelmeyer (2004) study in the full
model. This allows us to adapt parts of their proofs to the MAR model
considered here. To begin, we will require the Fisher information for
location of the error distribution to be finite:
\begin{assumption} \label{assumpF}
The error density $f$ is absolutely continuous with almost everywhere
derivative $f'$ satisfying
\begin{equation*}
J = \int_{-\infty}^{\infty} \ell^2(z)f(z)\,dz < \infty,
\end{equation*}
where $J$ is the Fisher information for location and $\ell = -f'/f$ is
the score function.
\end{assumption}

We do not assume a parametric model for the regression function or for
the distribution of the observations. The parameter set $\Theta$ of
the statistical model therefore includes a family of covariate
distributions $\GG$ satisfying Assumption \ref{assumpG}, a family of
error distributions $\FF$ satisfying Assumption \ref{assumpF}, a space
of regression functions $\RR$ that belong to $H(d,\gamma)$, and a
family of response probability distributions $\BB$ that are
characterized by proportion functions mapping $[0,\,1]^m$ to
$(0,\,1]$. It follows that we can write $\Theta = \GG \times \FF
\times \RR \times \BB$.

Since the construction of the efficient influence function utilizes
the directional information in $\Theta$, we will now identify the set
$\dot \Theta$ of all perturbations related to the statistical model,
which may be thought of as directions. The joint distribution
$P(dx,dy,dz)$ depends on the marginal distribution $G(dx)$ of $X$, the
conditional probability $\pi(x)$ that $\delta$ equals one given $X=x$,
and the conditional distribution $Q(dy\,|\,x)$ of $Y$ given $X=x$:
\begin{equation*}
P(dx,dy,dz) = G(dx)B_{\pi(x)}(dz\,|\,x)\Big\{
 zQ(dy\,|\,x) + (1 - z)\delta_0(dy)\Big\},
\end{equation*} 
where $B_p = p\delta _1 + (1 - p)\delta_0$ denotes the Bernoulli
distribution with parameter $p$ and $\delta_t$ is the Dirac measure
for $\{t\}$.

Now consider perturbations $G_{nu}$, $\pi_{nw}$ and $Q_{nv}$ of $G$,
$\pi$ and $Q$, respectively, that are {\em Hellinger differentiable}
in the following sense:
\begin{align*}
\int_{[0,\,1]^m} \bigg\{
 n^{1/2}\Big\{dG_{nu}^{1/2}(x) - dG^{1/2}(x)\Big\}
 &- \frac12 u(x)dG^{1/2}(x)\bigg\}^2\,dx
 \rightarrow 0, \\
\int_{[0,\,1]^m} \int_{\{0,\,1\}} \bigg\{
 n^{1/2}\Big\{dB_{\pi_{nw}}^{1/2}(z\,|\,x) - dB_{\pi}^{1/2}(z\,|\,x)\Big\}
 &- \frac12 \{z - \pi(x)\}w(x)dB_{\pi}^{1/2}(z\,|\,x)\bigg\}^2\,G(dx)
 \rightarrow 0, \\ 
\int_{[0,\,1]^m} \int_{-\infty}^{\infty} \bigg\{
 n^{1/2}\Big\{dQ_{nv}^{1/2}(y\,|\,x) - dQ^{1/2}(y\,|\,x)\Big\}
 &- \frac12 v(x,y)dQ^{1/2}(y\,|\,x)\bigg\}^2\,G_1(dx)
 \rightarrow 0, 
\end{align*} 
writing $G_1$ for the conditional distribution of $X$ given that
$\delta =1$. The perturbed distribution functions $G_{nu}$,
$B_{\pi_{nw}}$ and $Q_{nv}$ must satisfy the original model
constraints, which requires their Hellinger derivatives to be
restricted to suitable function spaces: $u$ belongs to
$\Lspace_{2,0}(G)$, i.e.\ $u \in \Lspace_2(G)$ and $\int_{[0,\,1]^m}
u(x)\,G(dx) = 0$; $w$ belongs to
\begin{equation*}
\Lspace_{2}(G_{\pi}) = \bigg\{ w \in \Lspace_2(G) \,:\,
 \int_{[0,\,1]^m} w^2(x) \pi(x)\{1 - \pi(x)\}\,G(dx) < \infty
 \bigg\},
\end{equation*}
writing $G_\pi(dx) = \pi(x)\{1 - \pi(x)\}G(dx)$; and $v$ belongs to
\begin{equation*}
\Vspace_0 = \bigg\{ v \in \Lspace_2(Q \otimes G_1) \,:\,
 \int_{-\infty}^{\infty} v(x,y)\,Q(dy\,|\,x) = 0 \bigg\}.
\end{equation*}

Note that models for $G_1$, $\pi$ and $Q$ will imply further
restrictions on the perturbations in order to satisfy those model
assumptions. This means that $u$, $w$ and $v$ must be further
restricted to subspaces of $\Lspace_{2,0}(G)$, $\Lspace_2(G_{\pi})$
and $\Vspace_0$, respectively. Here no model assumptions on $G$ and
$\pi$ have been introduced, but model equation \eqref{modeleq} does
present a structural constraint on the conditional distribution $Q$
of $Y$ given $X$. This implies that we only have to identify the
appropriate subspace $\Vspace$ of $\Vspace_0$ to account for this
additional structure. 

Since the covariates and the errors are assumed to be independent, we
may write the density function $dQ$ of $Q$ as $dQ(x,y) = f(y -
r(x))$. Using this notation, the constraint on $v \in \mathcal{V}_0$
now states that
\begin{equation*}
\int_{-\infty}^{\infty} v(x,y)f(y - r(x))dy = 0.
\end{equation*}
In order to derive the explicit form of the function space
$\mathcal{V}$, we introduce further respective perturbations $s$ and
$t$ for the unknown functions $f$ and $r$, and we can write
\begin{equation*}
dQ_{nv}(x,y) = dQ_{nst}(x,y) = f_{ns}\big(y - r_{nt}(x)\big),
\end{equation*}
where $f_{ns}(z) = f(z)\{1 + n^{-1/2}s(z)\}$ and $r_{nt}(x) = r(x) +
n^{-1/2}t(x)$ with $s \in \mathcal{S}$ and $t \in \Lspace_2(G_1)$. Our
assumptions on model \eqref{modeleq} require the errors to have mean
zero and the perturbed error density $f_{ns}$ must integrate to
one. Hence, $\Sspace$ takes the form
\begin{equation*}
\Sspace = \bigg\{s \in \Lspace_2(F) \,:\,
 \int_{-\infty}^{\infty} s(z)f(z)\,dz = 0,
 \int_{-\infty}^{\infty} zs(z)f(z)\,dz = 0 \bigg\}.
\end{equation*}
We can simply restrict the perturbation $t$ to belong to
$\Lspace_2(G_1)$, which follows from the fact that we do not assume a
parametric form for $r$.

With the appropriate spaces $\Sspace$ and $\Lspace_2(G_1)$ identified,
we can specify the appropriate form of $\Vspace$. In the following
arguments we will write ``$\doteq$'' to denote asymptotic equivalence,
i.e.\ equality up to an additive term of order $o_p(n^{-1/2})$. As in
M\"uller (2009), who considers a {\em parametric} (nonlinear)
regression function, a brief sketch gives
\begin{align*} 
f_{ns}\big(y - r_{nt}(x)\big)
 &= f\big(y - r_{nt}(x)\big)
 \Big\{1 + n^{-1/2}s\big(y - r_{nt}(x)\big)\Big\} \\
 &= f\big(y - r(x) - n^{-1/2}t(x)\big)
 \Big\{1 + n^{-1/2}s\big(y - r(x) - n^{-1/2}t(x)\big)\Big\} \\
 &\doteq f(y - r(x))
 \Big\{1 + n^{-1/2}\big\{s(y - r(x)) + \ell(y - r(x))t(x)\big\}\Big\}.
\end{align*}
Hence, we can write
\begin{equation} \label{Qperturb}
dQ_{nst}(x,y) \doteq f(y - r(x))
 \Big\{1 + n^{-1/2}\big\{s(y - r(x)) + \ell(y - r(x))t(x)\big\}\Big\}
\end{equation}
Equation \eqref{Qperturb} implies that $\Vspace$ has the form
\begin{equation*}
\Vspace = \bigg\{ v(x,y) = s(y - r(x)) + \ell(y - r(x))t(x) \,:\,
 s \in \Sspace \text{ and } t \in \Lspace_2(G_1) \bigg\}.
\end{equation*}

We can see that $\dot \Theta$ is the set containing all possible
Hellinger perturbations of the statistical model parameters: 
\begin{equation*}
\dot \Theta = \Lspace_{2,0}(G) \times \Sspace \times \Lspace_2(G_1)
 \times \Lspace_2(G_\pi).
\end{equation*}
The perturbed distribution $P_{n\gamma}$, with $\gamma = (u,s,t,w)$ in
$\dot \Theta$, of the observation $(X,\delta Y, \delta)$ can be
written
\begin{equation*}
P_{n\gamma}(dx,dy,dz) \doteq G_{nu}(dx)B_{\pi_{nw(x)}}(dz\,|\,x)
 \Big\{zQ_{nst}(dy\,|\,x) + (1 - z)\delta_0(dy)\bigg\}.
\end{equation*}
It then follows that $P_{n\gamma}$ is Hellinger differentiable with
perturbation function
\begin{equation} \label{Pperturb}
d_{\gamma}(x, y, z)
 = u(x) + z\big\{s(y - r(x)) + \ell(y - r(x))t(x)\big\}
 + \{z -\pi(x)\}w(x),
\end{equation}
and we have the stochastic expansion, writing $dP$ for the density
function of $P$,
\begin{equation*}
\sj \log\bigg( \frac{dP_{n\gamma}(X_j, \delta_j Y_j, \delta_j)}{
 dP(X_j, \delta_j Y_j, \delta_j)} \bigg)
 = n^{-1/2} \sj d_{\gamma}(X_j, \delta_j Y_j, \delta_j)
 - \frac12 E\big[d_{\gamma}^2(X, \delta Y, \delta)\big]
 + \op.
\end{equation*}
Since $n^{-1/2} \sj d_{\gamma}(X_j, \delta_j Y_j, \delta_j)$ is
asymptotically normally distributed with mean zero and variance
$E[d_{\gamma}^2(X, \delta Y, \delta)]$ by the central limit theorem,
it follows for the expansion above to characterize local asymptotic
normality in the present situation.

The efficient influence function of a differentiable functional is
characterized by its canonical gradient, which takes the form
$d_{\gamma}^*(X, \delta Y, \delta)$ for some $\gamma^* \in \dot
\Theta$. This gradient is defined as the orthogonal projection of the
gradient for the functional $E[h(\ve)]$ (to be specified later) onto
the tangent space given by the perturbed distributions
$P_{n\gamma}$. It then follows from \eqref{Pperturb} for the tangent
space $\Tspace$ to be equal to the closure of the linear subspace
formed by $d_{\gamma}$. Since $d_{\gamma}$ is a sum of orthogonal
elements we can write
\begin{equation*}
\Tspace = \Big\{u(X)\,:\,u \in \Lspace_{2,0}(G)\Big\}
 \oplus \Big\{\delta v(X,Y)\,:\, v \in \Vspace\Big\}
 \oplus \Big\{ \{\delta - \pi(X)\}w(X) \,:\,
 w \in \Lspace_2(G_\pi) \Big\}.
\end{equation*}

We are interested in the linear functional $E[h(\ve)]$. In order to
specify a gradient for $E[h(\ve)]$, we need the directional derivative
$\gamma_h \in \dot \Theta$ of $E[h(\ve)]$, which is characterized by a
limit as follows. As in M\"uller, Schick and Wefelmeyer (2004) we
have, for every $s \in \mathcal S$, 
\begin{equation*}
\lim_{n \to \infty} n^{1/2}\bigg\{
 \int_{-\infty}^{\infty} h(z)f_{ns}(z)\,dz
 - \int_{-\infty}^{\infty} h(z)f(z)\,dz \bigg\}
 = E\big[h(\ve)s(\ve)\big] = E\big[h_0(\ve)s(\ve)\big],
\end{equation*}
where $h_0$ is the projection of $h$ onto $\mathcal{S}$:
\begin{equation} \label{EhDeriv}
h_0(z) = h(z) - E\big[h(\ve)\big]
 - \frac{z}{\sigma^2}E\big[\ve h(\ve)\big].
\end{equation}
Here $\sigma^2$ denotes the error variance. Hence, $E[h(\ve)]$ is
directionally differentiable, and \eqref{EhDeriv} implies this
directional derivative is $\gamma_h = (0,h_0,0,0)$. It then follows
for $E[h(\ve)]$ to have the gradient $h_0(\ve)$, with $h_0$ given by
\eqref{EhDeriv}.

By the convolution theorem (see, for example, Section 2 of Schick,
1993), the unique canonical gradient $g^*(X, \delta  Y, \delta)$ is
obtained by orthogonally projecting the gradient $h_0(\ve)$ of
$E[h(\ve)]$ onto the tangent space $\Tspace$. Hence, $g^*(X, \delta Y,
\delta)$ must be of the form
\begin{equation} \label{gstar}
g^*(X, \delta Y, \delta) =
 u^*(X) + \delta\big\{s^*(\ve) + \ell(\ve)t^*(X)\big\}
 + \big\{\delta - \pi(X)\big\}w^*(X),
\end{equation}
which satisfies
\begin{align} \label{OrthProj}
E\big[h_0(\ve)s(\ve)\big] = E\big[ g^*(X, \delta Y, \delta)
 d_\gamma(X, \delta  Y, \delta) \big]
\end{align} 
for every $\gamma \in \dot \Theta$. A straightforward calculation
shows the right-hand side of \eqref{OrthProj} is equal to
\begin{align*}
&E\big[ g^*(X, \delta  Y, \delta)
 d_{\gamma}(X, \delta  Y, \delta) \big] \\
&= E\big[u^*(X)u(X)\big]
 + E\big[ \delta\big\{s^*(\ve) + \ell(\ve)t(X)\big\}
 \big\{s(\ve) + \ell(\ve)t^*(X)\big\} \big] \\
&\quad + E\big[\{\delta - \pi(X)\}^2w^*(X)w(X)\big] \\
&= E\big[u^*(X)u(X)\big] + E\delta \Big\{
 E\big[s^*(\ve)s(\ve)\big]
 + E\big[\ell_0(\ve)s^*(\ve)\big]E_1\big[t(X)\big] \\
&\quad + E\big[\ell_0(\ve)s(\ve)\big]E_1\big[t^*(X)\big]
 + JE_1\big[t^*(X)t(X)\big] \Big\}
 + E\big[\pi(X)\{1 - \pi(X)\}w^*(X)w(X)\big],
\end{align*} 
where $J$ is the Fisher information given in Assumption \ref{assumpF}
and $\ell_0(\ve)$ is the projection of $\ell(\ve)$ onto $\mathcal{V}$,
i.e.\ $\ell_0(\ve) = \ell(\ve) - (\ve/\sigma^2)$. The notation $E_1$
indicates the expectation is with respect to the conditional
distribution $G_1$. For convenience, we introduce the quantity $J_0$
which is calculated analogously to $J$:
\begin{equation*}
J_0 = E\big[\ell_0^2(\ve)\big]
 = E\Bigg[\bigg\{\ell(\ve) - \frac{\ve}{\sigma^2}\bigg\}^{2}\Bigg]
 = J - \frac{1}{\sigma^{2}}.
\end{equation*}

With appropriate choices of $\gamma$ in $\dot \Theta$, it easily
follows from \eqref{OrthProj} for $u^* = w^* = 0$. Now choosing the zero
function for the function $u$, \eqref{OrthProj} becomes
\begin{align*}
E\big[h_0(\ve)s(\ve)\big] = E\delta \Big\{
 E\big[s(\ve)s^*(\ve)\big]
 + E\big[\ell_0(\ve)s(\ve)\big]E_1\big[t^*(X)\big] \Big\},
\end{align*} 
which must hold for all $s \in \Sspace$. This implies
\begin{equation} \label{sstar}
s^*(z) = \frac{h_0(z)}{E\delta} - \ell_0(z)E_1\big[t^*(X)\big],
 \qquad z \in \R.
\end{equation}

Following M\"uller, Schick and Wefelmeyer (2004), we can consider
$\mathcal L_2(G_1)$ written as an orthogonal sum of functions with
mean zero and of constants, i.e.\ we write $\mathcal L_2(G_1) =
\mathcal L_{2,0}(G_1) \oplus [1]$. This means we can decompose $t(X)$
into $t(X) = \{t(X) - E_1[t(X)]\} + E_1[t(X)]$. Finally, choosing the
zero function for $s$ and inserting $s^*$ from \eqref{sstar},
\eqref{OrthProj} becomes
\begin{align*}
0 &= \frac1{E\delta}E\big[\ell_0(\ve)h_0(\ve)\big]E_1\big[t(X)\big]
 + (J - J_0)E_1\big[t^*(X)\big]E_1\big[t(X)\big] \\
&\quad + JE_1\bigg[ \Big\{t^*(X) - E_1\big[t^*(X)\big]\Big\}
 \Big\{t(X) - E_1\big[t(X)\big]\Big\} \bigg] \\
&= \bigg\{ \frac1{E\delta}E\big[\ell_0(\ve)h_0(\ve)\big]
 + \frac1{\sigma^2}E_1\big[t^*(X)\big] \bigg\}E_1\big[t(X)\big] \\
&\quad +  JE_1\bigg[ \Big\{t^*(X) - E_1\big[t^*(X)\big]\Big\}
 \Big\{t(X) - E_1\big[t(X)\big]\Big\} \bigg],
\end{align*}
which must hold for every $t \in \mathcal L_2(G_1)$. This implies
$t^*$ is equal to its mean $E_1[t^*(X)]$ and
\begin{equation} \label{tstar}
E_1\big[t^*(X)\big] = -\frac{\sigma^2}{E\delta}
 E\big[h_0(\ve)\ell_0(\ve)\big].
\end{equation} 
Therefore, combining \eqref{tstar} with the fact that $t^*$ must be
equal to its mean yields $t^*(x) =
-\sigma^2(E\delta)^{-1}E[h_0(\ve)\ell_0(\ve)]$, $x \in
[0,\,1]^m$. Combining the fact that $s^* = w^* = 0$ with \eqref{sstar}
and \eqref{tstar}, we obtain the following result: 
\begin{lemma} \label{lemCanGradEh}
The canonical gradient of $E[h(\ve)]$ is $g^*(X, \delta Y, \delta)$
given in \eqref{gstar} that characterized by $(0,s^*,t^*,0)$, with
\begin{gather*}
s^*(z) = \frac{1}{E\delta}\Big\{ 
 h_0(z) + \sigma^2E\big[h_0(\ve)\ell_0(\ve)\big]\ell_0(z)\Big\},
 \qquad z \in \R, \\
t^*(x) = -\frac{\sigma^2}{E\delta}E\big[h_0(\ve)\ell_0(\ve)\big],
 \qquad x \in [0,\,1]^m,
\end{gather*}
where $\sigma^2 = E[\ve^2]$ is the error variance, $h_0$ is given in
\eqref{EhDeriv} and $\ell_0(\ve) = \ell(\ve) - (\ve/\sigma^2)$.
\end{lemma} 

An estimator $\hat{\mu}$ of $E[h(\ve)]$ is called {\em efficient}, in
the sense of H\'ajek and Le Cam, when $\hat{\mu}$ is asymptotically
linear with influence function equal to the canonical gradient $g^*(X,
\delta Y, \delta)$ that characterizes $E[h(\ve)]$, i.e.\ if the
expansion holds:
\begin{equation*}
n^{1/2}\Big\{\hat{\mu} - E\big[h(\ve)\big]\Big\}
 = n^{-1/2}\sj g^*(X_j, \delta_j Y_j, \delta_j) + o_p(1).
\end{equation*}
A straightforward calculation combining the result of Lemma
\ref{lemCanGradEh} with the display above and formula \eqref{gstar} yields:
\begin{corollary} \label{corEhEffIF}
Consider the nonparametric regression model with responses missing at
random. An efficient estimator $\hat{\mu}$ of $E[h(\ve)]$ must satisfy
the expansion
\begin{equation*}
n^{1/2}\Big\{\hat{\mu} - E\big[h(\ve)\big]\Big\}
 = n^{-1/2}\sj \frac{\delta_j}{E\delta} \Big\{
 h(\ve_j) - E\big[h(\ve)\big] - E\big[\ell(\ve)h(\ve)\big]\ve_j \Big\}
 + \op.
\end{equation*}
\end{corollary}

\begin{remark} \label{remMSWestimates}
M\"uller, Schick and Wefelmeyer (2004) construct residual-based
estimators $n^{-1} \sj h(\hve_j)$ for estimating $E[h(\ve)]$ in the
full model. In their Section 2, they give conditions for the i.i.d.\
representation:
\begin{equation*}
n^{-1/2} \sj h(\hve_j) = n^{-1/2} \sj
 \Big\{h(\ve_j) - E\big[h'(\ve)\big]\ve_j\Big\} + \op,
\end{equation*}
which characterizes an efficient estimator. (For simplicity, we assume
in this remark that $h$ is differentiable.) Note that $E[h'(\ve)] =
E[\ell(\ve)h(\ve)]$. Hence, using the transfer principle, we see
that the complete case versions of their estimators have the expansion
from the previous corollary, and, therefore, these estimates are also
efficient in the MAR model.
\end{remark}

The function $h(\ve) = \1[\ve \leq t]$ is of particular interest
because many statistical methods are residual-based and require
estimation of the error distribution function. Using Corollary
\ref{corEhEffIF} with this particular $h(\ve)$, we obtain an expansion
for the residual-based empirical distribution function:
\begin{corollary} \label{corFhatEffExpan}
Consider the nonparametric regression model with responses missing at
random. An estimator $\hat F$ of the error distribution function $F$
is efficient, if it satisfies the expansion
\begin{equation*}
n^{1/2}\Big\{\hat{F}(t) - F(t)\Big\}
 = n^{-1/2} \sj \frac{\delta_j}{E\delta}
 \Big\{\1\big[\ve_j \leq t\big] - F(t) + f(t)\ve_j\Big\} + \op.
\end{equation*}
\end{corollary}
Note, this is the expansion of the complete case estimator $\Fhat_c$
from the previous section, which provide the proof of the second
assertion in Theorem \ref{thmFhatCCExpan}.


\section{Simulation results} \label{sims}
To conclude the article, we present a brief simulation study of the
previous results in two important examples. The first example
compares the efficiency property of $\Fhat_c$, which is constructed
using only the completely observed data, to another estimator
$\Ftilde$, which is constructed using an imputation methodology. In the
second example, we consider applying a goodness-of-fit test for
normal errors to the residuals from the nonparametric regression
constructed by only the complete cases and by the same imputation
methodology that was implemented in the first example. In both
examples, we assume a nonparametric regression model \eqref{modeleq},
but choosing
\begin{equation*}
r(x) = x^3 - x^2 + x + \cos\big((3\pi/2)x\big),
\end{equation*}
which we expect preserves the nonparametric nature of the studies.

\begin{figure} 
\centering
\includegraphics[height=80mm]{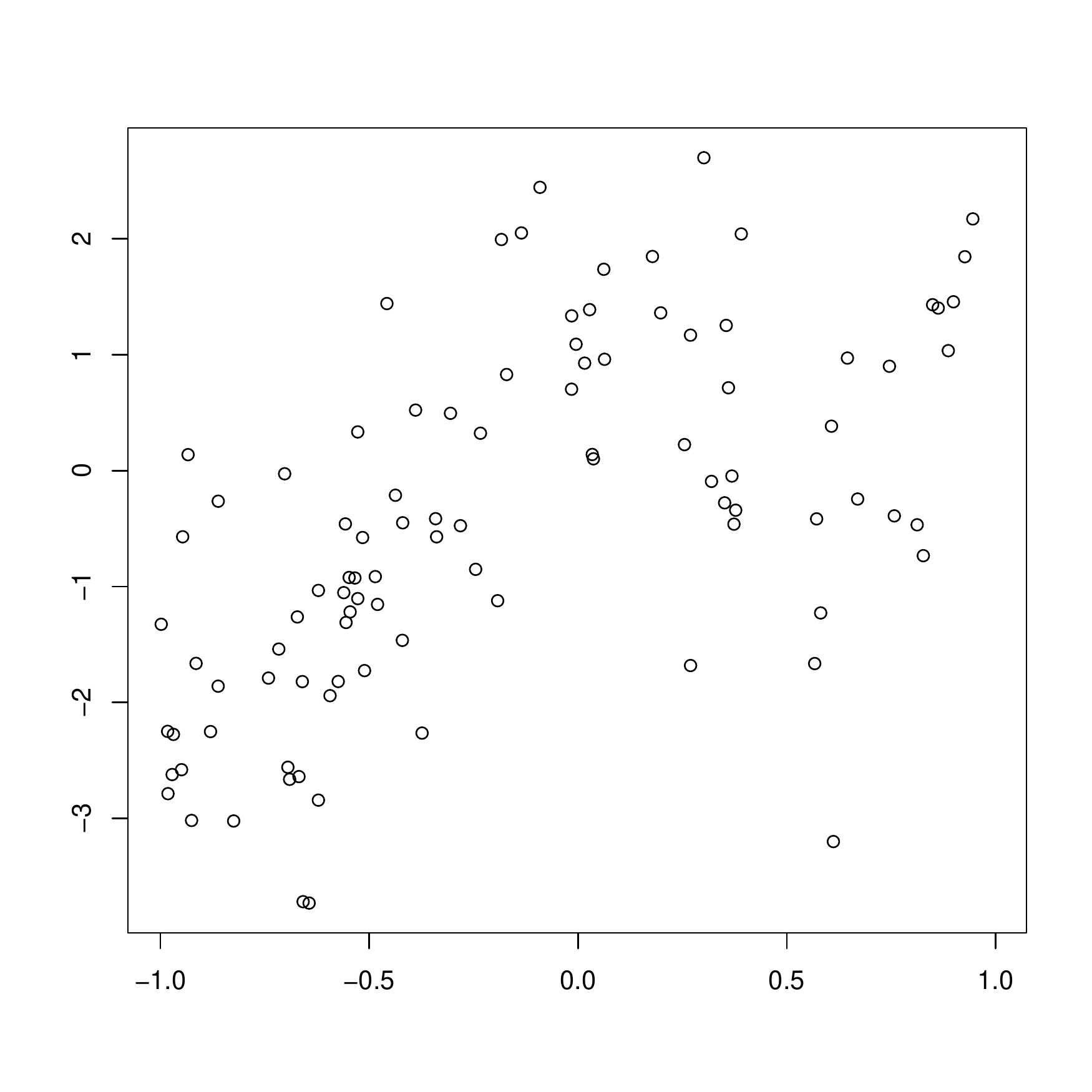} 
\vspace{1ex}
\caption{{\it A scatter plot of $100$ data points from a typical
    simulated dataset.}}
\label{figure1}
\end{figure} 

The covariates are generated from a uniform distribution and the
errors from a normal distribution: $X_j \sim U(-1,1)$ and $\ve_j \sim
N(0,1)$ for $j=1,\dots,n$; see Figure \ref{figure1} for a scatterplot
of a typical simulated dataset. Finally, the indicators $\delta_j$
have a Bernoulli distribution with proportion function parameter
$\pi(x) = P(\delta=1|X=x)$, which is chosen to be the logistic
distribution function with a mean of zero and scale parameter of one:
\begin{equation*}
\pi(x) \, = \, \frac{1}{1+e^{-x}}.
\end{equation*}
Consequently, the average amount of missing data is 50\% and ranges
between 27\% and 73\%. Finally, we work with the local linear smoother
(it is easy to see that $r$ is Lipschitz and, therefore, it belongs to
the H\"older space $H(1,1)$), and the bandwidth sequence $\{c_n\}_{n
  \geq 1}$ is taken as $c_n = 1.25 \{n \log(n)\}^{-1/4}$. The
assumptions of Theorem \ref{thmFhatCCExpan} are then satisfied.


\subsection{Example 1: Simulation of asymptotic mean squared error}
We consider two estimators of the error distribution function. The
first estimator is the proposed complete case estimator $\Fhat_c$ and
the second is a `tuned' version of $\Fhat_c$ that utilizes an
imputation technique. Similar to Gonz\'alez-Manteiga and
P\'erez-Gonz\'alez (2006), we take the initial local polynomial
complete case estimator $\rhat_c$ (see equation \eqref{locpolyCC}) to
produce the completed sample $(X_j,\hat Y_j)$, $j = 1, \dots, n$. We
chose to fully impute the responses, i.e.\ $\hat Y_j =
\rhat_c(X_j)$. This is a variation of the approach of
Gonz\'alez-Manteiga and P\'erez-Gonz\'alez (2006), who work with partially
imputed responses $\hat Y_j = \delta_j Y_j + (1 -
\delta_j)\rhat_c(X_i)$. A new local polynomial estimator
$\rhat^*(\cdot)$ is then constructed from the completed sample. When
$Y$ is observed, we can compute adjusted residuals $\hat \ve_j^* = Y_j
- \hat r^*(X_j)$ based on the updated estimator $\rhat^*$. This means
we still work with complete cases when estimating $F$, but now the
estimated regression has changed.

Using these residuals we obtain the new tuned estimator 
\begin{equation*}
\Ftilde(t) = \frac1N \sj \delta_j\1\big[\hve_j^* \leq t\big].
\end{equation*}
The results in the previous sections show the complete case estimator
$\Fhat_c$ is an (asymptotically) efficient estimator of $F$. Our
discussion in Remark \ref{remNoTTPrin} also suggests the tuned
estimator $\Ftilde$ is also efficient; i.e.\ both estimators are
asymptotically equivalent. We expect that $\Ftilde$ can be expanded in
the same way as $\Fhat_c$:
\begin{equation*}
\sup_{t \in \R} \bigg| \avj \frac{\delta_j}{E\delta} \Big\{
 \1\big[\hve_j^* \leq t\big] - \1\big[\ve_j \leq t\big]\Big\} -
 f(t) \int_{[0,\,1]^m} \ahat^*(x)\,G_1(dx) \bigg|
 = \opn,
\end{equation*} 
where $\ahat^*(x)$ is now an approximation of the difference
$\rhat^*(x) - r(x)$ (cf.\ equation (\ref{approx}) in Remark
\ref{remNoTTPrin}). The integral in the display above can be written
as
\begin{equation*}
\int_{[0,\,1]^m} \ahat^*(x)\,G_1(dx)
 = \int_{[0,\,1]^m} \ahat_c(x)\,G_1(dx)
 + \int_{[0,\,1]^m} \big\{\ahat^*(x) - \ahat_c(x)\big\}\,G_1(dx).
\end{equation*}
Since $\ahat^*(x) - \ahat_c(x)$ approximates the difference
$\rhat^*(x) - \rhat_c(x)$ of two consistent estimators of $r(x)$, we
expect the last term in the display above to be asymptotically
negligible. Repeating the arguments from Remark \ref{remNoTTPrin}
would then give the desired expansion:
\begin{equation*}
\sup_{t \in \R} \bigg| \frac1N \sj \delta _j \Big\{
 \1\big[\hve_j^* \leq t\big] - \1\big[\ve_j \leq t\big]
 - \ve_j f(t) \Big\} \bigg|
 = \opn,
\end{equation*}
and, hence, both $\Fhat_c$ and $\Ftilde$ should have the same
asymptotic expansion, i.e.\ both estimators are asymptotically
equivalent.

In order to further check the conjecture that both estimators are
asymptotically equivalent, we conducted a simulation study using 1000
trials. We considered four sample sizes and five different points at
which the error distribution function was evaluated. The findings are
summarized in Table \ref{table1}. Note, we also implemented another
estimator, which uses partial imputation to complete the sample as
suggested by Gonz\'alez-Manteiga and P\'erez-Gonz\'alez (2006), but
our approach performed slightly better and so we only report the
results for our tuned estimator $\Ftilde$. For the second smoothing
step we chose the same bandwidth as in the first step, $c_n = 1.25 \{n
\log(n)\}^{-1/4}$.

\begin{table}
\centering
\begin{tabular}{|r|c c|c c|c c|c c|c c|}
\hline
\diagbox{$n$}{$t$} & \multicolumn{2}{|c|}{$-1.5$} &
 \multicolumn{2}{|c|}{$-1$} & \multicolumn{2}{|c|}{$0$} &
 \multicolumn{2}{|c|}{$1$} & \multicolumn{2}{|c|}{$1.5$} \\
\hline
50 & 0.1141 & 0.0987 & 0.2705 & 0.2087 & 0.1702 & 0.1884 & 0.2865 &
 0.2220 & 0.1179 & 0.1009 \\
250 & 0.1018 & 0.0930 & 0.1800 & 0.1634 & 0.2021 & 0.2071 & 0.2022 &
 0.1972 & 0.1201 & 0.1165 \\
1000 & 0.0991 & 0.0945 & 0.1668 & 0.1625 & 0.1865 & 0.1997 & 0.1706 &
 0.1780 & 0.1000 & 0.1008 \\
10000 & 0.0925 & 0.0920 & 0.1567 & 0.1537 & 0.2068 & 0.2274 & 0.1690 &
 0.1752 & 0.0953 & 0.0975 \\
true & 0.0911 & -- & 0.1498 & -- & 0.1816 & -- & 0.1498 & -- & 0.0911
 & -- \\
\hline
\end{tabular}
\vspace{1ex}
\caption{{\it Simulated and true asymptotic MSE of $n^{1/2}\{\Fhat_c -
    F\}$ and $n^{1/2}\{\Ftilde - F\}$ at the points $-1.5$, $-1$, $0$,
    $1$ and $1.5$.}}
\label{table1}
\end{table}

These results show the simulated MSE (multiplied by $n$) of the
efficient estimator $\Fhat_c$ is very close to the true asymptotic MSE
(which equals the asymptotic variance and can be calculated using the
results of Theorem \ref{thmFhatCCExpan}). We can also see the
asymptotic MSE estimates of $\Ftilde$ are similar to those of
$\Fhat_c$ at large sample sizes. This provides further evidence for
the two approaches to be asymptotically equivalent as conjectured. The
simulated MSE values of $\Ftilde$, however, more closely match the
true asymptotic MSE values at low sample sizes, which we expect is due
to the imputation technique. However, at the point $0$ both estimators
perform very similarly for all sample sizes. A possible explanation of
this behavior is the point $0$ is also the median of this error
distribution, and we believe the imputation technique is least helpful
in this case from our discussion above conjecturing on the expansion
of $\Ftilde$.


\subsection{Example 2: Simulating a goodness-of-fit test for normal
  errors}
We now consider a test proposed by M\"uller, Schick and Wefelmeyer
(2012) for the full model with multivariate covariates. This test was
also examined by Koul, M\"uller and Schick (2012) in the MAR model
with a one-dimensional covariate, but without simulations. Both
articles study versions of a martingale transform test developed by
Khmaladze and Koul (2009). Under the null hypothesis, these test
statistics have limiting distributions given by $\sup_{0 < t \leq 1}
|B(t)|$, where $B(t)$ is the standard Brownian motion. These test
statistics are asymptotically distribution free because the limiting
distribution does not depend on any unknown parameters, which would
have to be estimated. This is very useful because the corresponding
complete case statistics have the same limiting distributions in this
case, which is a consequence of the transfer principle. Hence, the
decision rule remains unchanged in the MAR model. For example, setting
the level of the test to 0.05, we reject $H_0$ when the test statistic
exceeds 2.2414, which is the upper 5\% quantile of the distribution of
$\sup_{0 < t \leq 1} |B(t)|$.

Writing $\phi$ for the density function of the standard normal
distribution and $\sigma^2$ for the error variance, the null
hypothesis of normal errors is
\begin{equation*}
H_0:\, \exists\, \sigma > 0 \,:\, f(t)
 = \frac1\sigma \phi\bigg(\frac{t}{\sigma}\bigg),\quad t \in \R.
\end{equation*}
In order to introduce the test statistic $T_c$, define $h(x) =
(1,-\phi'(x)/\phi(x),-(x\phi(x))'/\phi(x))^T$ and 
\begin{equation*}
H(t) = \int_{-\infty}^t
 h^T(t)\Gamma^{-1}(t)\phi(t)\,dt,
\end{equation*}
where $\Gamma(t) = \int_t^{\infty} h(u)h^T(u)\phi(u)\,du$ (see
M\"uller, Schick and Wefelmeyer, 2012, and Koul, M\"uller and Schick,
2012, for an explicit definition of $\Gamma$ and for more
details). Following Koul, M\"uller and Schick (2012), we have the test
statistic
\begin{equation*}
T_c = \sup_{t \in \R} \bigg| N^{-1/2} \sj
 \delta_j\Big\{ \1\big[\hat Z_{j,c} \leq t\big]
 - H\big(t \wedge \hat Z_{j,c}\big) h\big(\hat Z_{j,c}\big)
  \Big\}\bigg|.
\end{equation*} 
Note, this statistic is based on our proposed estimator $\Fhat_c$, but
with {\em scaled} residuals $\hat Z_{j,c} = \hat \ve_{j,c}/\hat
\sigma_c$, where $\hat \sigma_c$ is the complete case version of the
residual-based empirical estimator, i.e.\ $\hat \sigma_c = \sqrt{\hat
  \sigma_c^2}$ with
\begin{equation*}
\hat \sigma_c^2 = \frac1N \sj \delta_j \hve_{j,c}^2
 = \frac 1N \sj \delta_j \big\{Y_j - \rhat_c(X_j)\big\}^2.
\end{equation*}
Recall that, under the MAR assumption, $\ve$ and $\delta$ are
independent. Hence, $\hat \sigma_c^2$ is a consistent estimator of
$\Var(\ve | \delta =1) = \Var(\ve) = \sigma^2$.

We are interested in studying the performance of $T_c$ in the MAR
model, and we wish to compare it with the corresponding statistic
$T_\iota$ that is based on the tuned estimator $\Ftilde$. Here
$T_\iota$ has exactly the same form as $T_c$ but with all $\hve_{j,c}$
replaced by the adjusted residuals $\hve_j^* = Y_j -
\rhat^*(X_j)$. For the simulations, we consider the same scenario as
in the previous example, but now also admit some other models for the
error distribution.

First we look at the $N(0,2)$ distribution to allow verification of
the ($5\%$) level of the test. To check the power of the test, we
generated errors from a mean shifted $\chi^2(1)$ distribution, a
$t(4)$ distribution and a $\text{Laplace}$ distribution with mean 0
and variance 2. The simulation study is based on 1000 runs and samples
of size 50 and 200.

\begin{table} 
\centering
\begin{tabular}{|r|c c|c c|c c|c c|}
\hline 
 & \multicolumn{2}{|c|}{ $N(0,2)$ } & \multicolumn{2}{|c|}{ $\chi^2_1
   - 1$ } & \multicolumn{2}{|c|}{ $t_4$ } & \multicolumn{2}{|c|}{
   Laplace$(0,2)$ } \\
\cline{2-9}
$n$ & $T_c$ & $T_\iota$  & $T_c$ & $T_\iota$  & $T_c$ & $T_\iota$ &
 $T_c$ & $T_\iota$  \\
\hline
50 & 0.022 & 0.025 & 0.489 & 0.535 & 0.099 & 0.108 & 0.095 & 0.119 \\
200 & 0.030 & 0.028 & 1.000 & 1.000 & 0.457 & 0.463 & 0.459 & 0.483 \\
\hline
\end{tabular}
\vspace{1ex}
\caption{Test for normally distributed errors. Simulated level is
  given by $N(0,2)$ figures.}
\label{table2}
\end{table}

Table \ref{table2} shows, when the errors are normally distributed
(and the null hypothesis is true), the test using $T_c$ rejects the
null hypothesis 2.2\% of the time for samples of size 50, and 3\% of
the time for samples of size 200. This indicates the test using $T_c$
is slightly conservative. We find similar conservative behavior in the
test using $T_\iota$, where the hypothesis of normality is rejected
2.5\% and 2.8\% of the time for sample sizes 50 and 200,
respectively. When the null hypothesis is not true, the power figures
are fairly close for both tests. The test using $T_\iota$ seems to be
more powerful for low sample sizes, which is expected from the results
of the first example. The differences are less pronounced for the
larger sample size of 200, suggesting that the two approaches are
asymptotically equivalent -- which is also what we would expect given
the discussion and the simulation results in the previous
example. Summing up, both test procedures have similar
performance. The test based on $T_c$ appears to be the better choice
for moderately large (or large) samples because it is easier to
implement.


\section*{Acknowledgements}
Ursula U. M\"uller was supported by NSF Grant DMS 0907014. The authors
thank the referees for a number of suggestions that improved the
manuscript. We also thank Susan Davis for her helpful comments on an
earlier draft.


\end{document}